\documentclass[numberedappendix,apj]{emulateapj}
\usepackage{times}
\usepackage{amssymb,amsmath,enumerate}
\usepackage{natbib}
\usepackage{graphicx}
\newcommand{\vv}{\mbox{\boldmath$v$}}

\newcommand{\SSS}{\mbox{\boldmath$S$}}

\shorttitle{Gravitational drag from a disk}
\shortauthors{Cant\'o et al.}
\slugcomment{Draft version, \today}

\begin{document}

\title{Gravitational drag on a point mass in hypersonic motion within
a Gaussian disk}

\author{
J. Cant\'o\altaffilmark{1},
A. Esquivel\altaffilmark{2},
F. J. S\'anchez-Salcedo\altaffilmark{1},
A. C. Raga\altaffilmark{2}}
\affil{
\altaffilmark{1}Instituto de Astronom\'\i a, Universidad
Nacional Aut\'onoma de  M\'exico, Ap. 70-468, 04510 D.F., M\'exico\\
\altaffilmark{2}Instituto de Ciencias Nucleares, Universidad Nacional
Aut\'{o}noma de M\'{e}xico, Apartado Postal 70-543, 04510 M\'{e}xico
D.F., M\'{e}xico\\
} 

\email{esquivel@nucleares.una.mx; jsanchez@astro.unam.mx; \\
raga@nucleares.unam.mx} 

\begin{abstract}
\label{sec:abstract}

We develop an analytical model for the accretion and gravitational
drag on a point mass that moves hypersonically in the midplane of a
gaseous disk with a Gaussian vertical density stratification.
Such a model is of interest for studying the interaction between a
planet and a protoplanetary disk, as well as the dynamical decay of massive
black holes in galactic nuclei.
The model considers that the flow is ballistic, and gives fully
analytical expressions for both the accretion rate onto the point mass,
and the gravitational drag it suffers. The expressions are further
simplified by taking the limits of a thick, and of a thin disk. The
results for the thick disk reduce correctly to those for a uniform density
environment (Cant\'o et al. 2011). 
We find that for a thin disk (small vertical scaleheight compared to the
gravitational radius) the accretion rate is proportional to the mass
of the moving object and to the surface density of the disk, while the
drag force is independent of the velocity of the object.  The
gravitational deceleration of the hypersonic perturber in a thin disk
was found to be independent of its parameters (i.e. mass or velocity)
and depends only on the surface mass density of the disk.
The predictions of the model are compared to the results of
three-dimensional hydrodynamical simulations, with a reasonable agreement.

\end{abstract}


\section{Introduction}

A gravitational point-like particle moving in a background gas can
capture ambient matter. In addition,
the particle may experience a drag due to the interaction with
the wake that it induces in the medium. Both accretion and dynamical friction
are ubiquitous phenomena in astrophysics, from protostars embedded
in molecular clumps to supermassive black holes at the center of galaxies
(e.g., S\'anchez-Salcedo 2012). 

Bondi, Hoyle and Lyttleton (Hoyle \& Lyttleton 1939, 1940a, 1940b, 1940c;
Bondi 1952) treated the problem by assuming
a point mass moving at constant speed within a uniform gaseous media. 
By neglecting the pressure of gas except on the downstream axis,
Bondi \& Hoyle (1944) derived a formula that gives the accretion rate as a 
function of the Mach number for highly supersonic flows. Later on,
Bondi (1952) suggested an approximate formula for any Mach number. 
Some authors have used Bondi's formula to account for the initial
mass function of stars as a competition for gas accretion by protostars
(Bonnell et al.~2001a,b; Klessen \& Burkert 2000, 2001)
and to estimate the rate at
which supermassive black holes accrete mass (e.g., Di Matteo et al.~2001).

The loss of momentum of the object relative to the cloud 
due to its gravitational interaction with its wake was also envisaged in
Bondi \& Hoyle (1944) using their line-accretion model.
Traditionally, however, the gravitational drag force has
been derived by calculating the density structure of the wake
in linear perturbation theory (Dokuchaev 1964; Ruderman
\& Spiegel 1971; Just \& Kegel 1990; Ostriker 1999; Kim \& Kim 2007;
S\'anchez-Salcedo 2009, 2012; Namouni 2010).
In this approach, a minimum radius $r_{\rm min}$ must be introduced
because the linear approximation is not valid close enough to the 
perturber. Recently, Cant\'o et al.~(2011) were able to calculate
the contribution of the nonlinear inner part of the wake to the
gravitational drag on hypersonic perturbers by using the ballistic
orbit theory. 

All the abovementioned papers have considered that the surrounding
gaseous medium is initially uniform, of constant density $\rho_{0}$. 
In this case and for a polytropic gas, the accretion rate and the 
gravitational drag are largely characterized by the Mach number.
Because of the assumption that the medium is infinite, the gravitational 
drag on supersonic bodies increases logarithmically in time 
(e.g., Ostriker 1999). In real life, however, systems are finite
in size.
Whilst the problem of the dynamical friction
in non-homogeneous (usually flattened) collisionless 
systems has received considerable attention
(eg., Binney 1977; Tremaine \& Weinberg 1984;  Maoz E. 1993; 
Colpi et al.~1999; Just \& Pe\~narrubia 2005), less studied   
is the case for gaseous systems. S\'anchez-Salcedo \& Brandenbrug (2001)
simulated the orbital decay of a satellite in orbit around a Plummer sphere
of gas and found
that the ``local approximation'', that is estimating the drag force at the
present location of the perturber as if the medium were homogeneous (but taking
appropriately the Coulomb logarithm) is very successful.

In this paper we study the gravitational drag on a body moving
in the midplane of a vertically-stratified gaseous disk.
It has been recently noted that the dynamical friction
in a gaseous slab could be useful to describe the interaction
between a planet in an eccentric orbit and the protoplanetary
disk (Muto et al.~2011).  These authors argue that 
the standard analysis of the interaction between a planet and
the disk, which assumes that the planet is corotating with the disk, is
not valid when the planet eccentricity exceeds the disk aspect ratio.
Instead, for moderate and large eccentricities, they suggest it is
more convenient to use the dynamical friction formula.
They derived the dynamical friction force felt by a perturber moving
in an infinitesimally thin disk of constant surface density and
calculated the migration and eccentricity damping timescales.
When the dynamical friction formula is applied to planets
with moderate eccentricities, the migration and eccentricity
damping timescales are in agreement with those found in previous
works. 

Gaseous dynamical friction is also relevant to study the 
disk-planet interaction in highly inclined systems. Rein (2012) 
computed the time-scales associated with migration and inclination
damping of planets due to dynamical friction with the protoplanetary
disk in the linear approximation and compared the results with
simulations. In order to model the gravitational interaction,
he uses a large softening radius, so that their simulations
also lie in the linear regime as well. This is justified as long
as the physical radius of the planet $r_{p}$ is much larger than
its accretion radius ($R_{A}\equiv 2GM/v^{2}$, where $v$ is the velocity
of planet relative to the gas in the disk).  For a hot Jupiter with
a mass of $100$ Earth masses and a size of $10$ Earth radii moving
on an inclined circular orbit around a star, the condition 
$r_{p}>R_{A}$ implies inclinations between the orbit of the planet
and the plane of the disk larger than $\sim 71^{\circ}$.
In the present work, we are interested in modelling the nonlinear
contribution of the gravitational wake to the drag force
on an accreting point-mass moving in non-inclined orbits.

A second area of application is in the orbital decay of 
massive black holes residing in galactic nuclei.
The formation of massive black hole binaries following galaxy
mergers is a natural consequence of the hierarchical scenario
(e.g., Begelman et al.~1980).
The gravitational torque due to dynamical friction with
an accretion disk formed by gas that was funnelled to the central $100$ pc,
 is expected to
be responsible for the in-spiral toward the center of the merger
remnant down to pc separations (Escala et al.~2005; Dotti et al.~2007;
Cuadra et al.~2009; Khan et al.~2011; Preto et al.~2011; Roedig et
al.~2012).
Recently, Nixon et al.~(2011a,b) showed that the binary orbital
angular momentum and energy are more efficiently removed in a
retrograde accretion disk than they are in a prograde
accretion disk because there are no orbital resonances between
the binary and the disk. Since there is no compelling reason
to assume that the binary and the disk rotation are initially
parallel, it is quite possible that retrograde disks strongly
promote coalescence of the black hole binary.

So far, most of the semi-analytical models aimed to study
the assembly of black holes use a prescription for the dynamical drag
(e.g., Tanaka \& Haiman 2009).
Given the importance of having analytical formulae to be applied
for planets or black holes in eccentric or counter-rotating orbits,
the strategy in our paper is to provide analytical estimates
for the gravitational drag on an accreting object embedded in a
three-dimensional disk of gas and to check its validity through 
numerical simulations. 
We will focus on highly supersonic perturbers. 
For instance, if a perturber is in orbit within a disk whose
aspect ratio is constant with radius, the relative
velocity between the perturber and its surroundings will be always
supersonic if its eccentricity exceeds the disk aspect ratio (see, for
instance, Muto et al. 2011).
Moreover, it is clear that perturbers in retrograde disks 
also move supersonically, even if the orbit is circular. 
We will concentrate on a Gaussian vertical profile, but the method is
general and can be applied for any decaying profile.

\label{sec:introduction}

\section{The free streaming model}
\label{sec:model}

\begin{figure}
  \centering
  \includegraphics[width=0.46\textwidth]{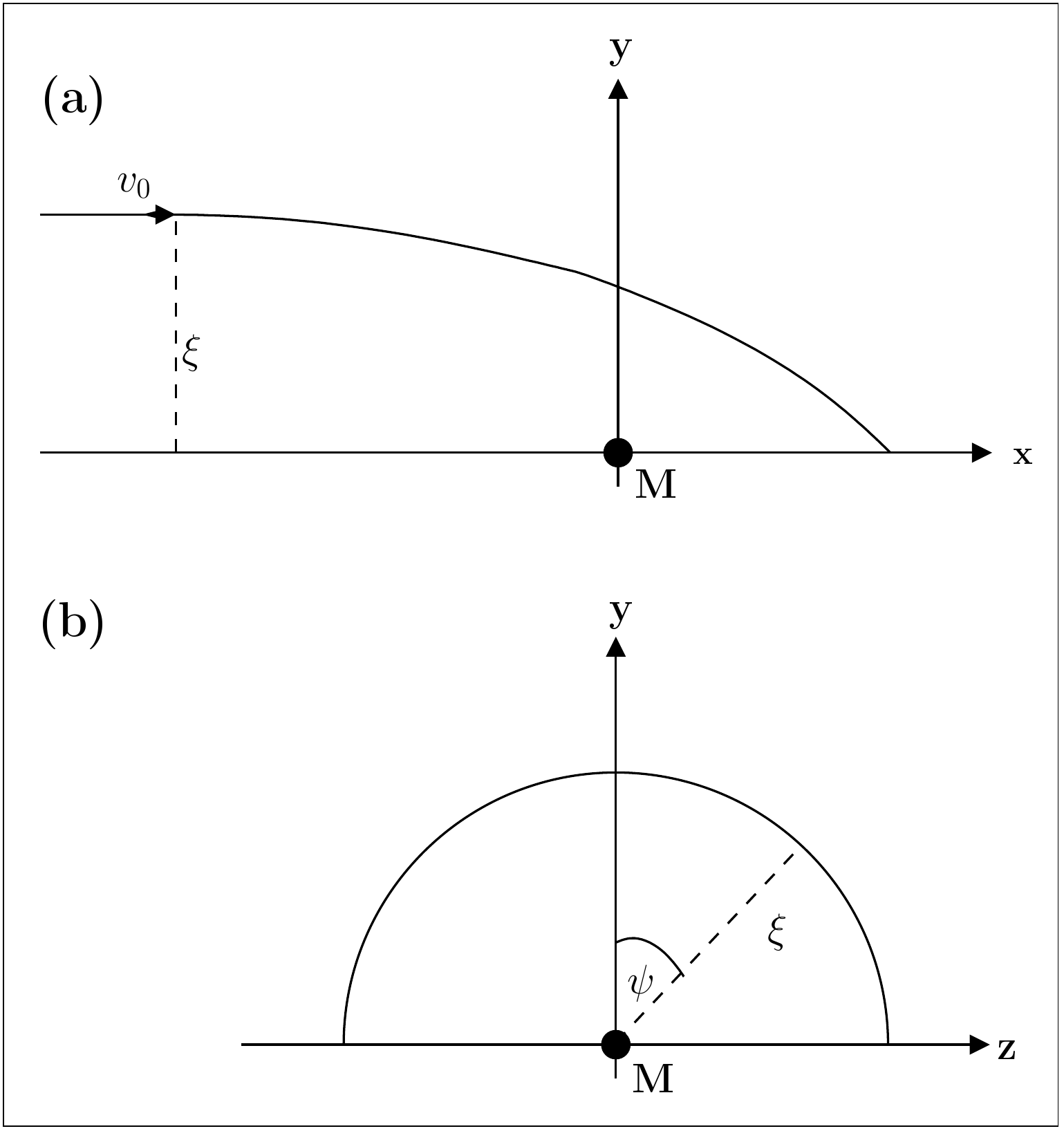}
  \caption{Schematic diagram of the geometry and reference frame used
    in the model (the mass $M$ is at rest). Far from the source a fluid
    parcel has a velocity $v_0$ (parallel to the $x$-axis),
    and an impact parameter $\xi$, but as
    it approaches the point mass its trajectory is deflected as
    illustrated in panel (a). In panel (b) the angle $\psi$ is
    shown in a view perpendicular to the relative motion of M
    with respect to the environment.}
  \label{fig:strat}
\end{figure}

We consider the problem of a point mass $M$ moving
hypersonically at velocity $v_{0}$ in a stratified, gaseous medium.
As a result of the gravitational pull produced by $M$ the
otherwise straight trajectory of a fluid parcel is bent, and a
wake behind the perturber is formed (see the top panel of Figure
\ref{fig:strat}).
In the hypersonic limit, the solution for the trajectory of a
fluid parcel (streamline) can be obtained neglecting the pressure
force and considering that the motion is ballistic.
By hypersonic we mean that the dynamical pressure is much larger
than the thermal pressure, i.e. $v_{0}^{2}\gg c_{s}^{2}$, where 
$c_{s}$ is the sound speed, so that the Mach number ${\mathcal{M}}$
satisfies ${\mathcal{M}}^2\gg 1$.
The case of a uniform environment was treated in \citet{2011MNRAS.418.1238C} 
and a comparison of the analytic model and a numerical simulation
was presented.

In the present paper we consider that the point mass moves in the
mid-plane of a plane-parallel structure, whose density decays
in the vertical direction $y$. For illustration, we will assume
a rather realistic Gaussian vertical density profile of the form
\begin{equation}
\rho\left(y\right)=\rho_0\,\mathrm{e}^{-y^2/h^2}=\rho_0\,\mathrm{e}^{-(\xi
  \cos\psi)^2/h^2},
\label{eq:strat}
\end{equation}
where $\rho_0$ is the midplane density, $h$ is the scaleheight, $\xi$ is
the fluid parcel impact parameter, and the angle $\psi$ is measured
from the $y$-axis (see Figure \ref{fig:strat}).
We adopt the same reference frame as in \citet{2011MNRAS.418.1238C},
in which the point mass $M$ is at rest. In this reference frame,
the motion can be modeled as a streaming environment with a constant
velocity $v_0$ far upstream of the source.
The undisturbed, streaming velocity $v_0$ lies
in the positive $x$ direction.

In a hypersonic regime, the trajectories
of the parcels, subject to the gravitational
field of $M$, have initially positive energy (per unit
mass) $E=v_o^2/2$ and therefore they are hyperbolic. 
The velocity of each parcel of gas is the same as in
\citet{2011MNRAS.418.1238C}, that is, the solution of the hyperbolic
trajectory after imposing the upstream $v_0$ boundary condition and
requiring angular momentum conservation along the streamlines.
Our model assumes that when the fluid encounters the $x$-axis,
the other velocity components ($v_y$ and $v_z$)
are thermalized at a shock, and the corresponding energy is
immediately radiated away. This condition results in a decrease of the
kinetic energy from $E_0=v_0^2/2$ at $x \rightarrow -\infty$ to
\begin{equation}
E_{\mathrm{tot}}=\frac{v_0^2}{2}\,
\left[1-\left(\frac{2\xi_0}{\xi}\right)^2\right],
\label{eq:Eax}
\end{equation}
where $\xi_0$ is the gravitational radius \citep[][]{1979SvA....23..201B,
  2011MNRAS.418.1238C}, defined as 
\begin{equation}
\xi_0\equiv \frac{GM}{v_0^2}\,.
\label{eq:xi0}
\end{equation}
From equation (\ref{eq:Eax}) it is clear that all the material with
an impact parameter $\xi\leq 2\xi_0$ after the shock has a negative
energy and thus remains gravitationally bound
(i.e., it will eventually be accreted onto the point mass).

From equation (\ref{eq:Eax}) we can also see that the material that is
gravitationally unbound when entering the wake will flow away along
the $x$-axis reaching infinity 
with a velocity
\begin{equation}
v_{\infty} = v_0\,\left[1-\left(\frac{2\xi_0}{\xi}\right)^2\right]^{1/2}.
\label{eq:vinf}
\end{equation}

\subsection{The mass accretion rate and the gravitational drag}

The accretion rate can be calculated by integrating over all the material
intercepted by a cylinder with an impact parameter up to $2\xi_0$:
\begin{equation}
\dot{M}_{\mathrm{acc}}=
\int_0^{2\xi_0}2\pi\,\bar{\rho}(\xi)\,v_0\,\xi\,\mathrm{d}\xi,
\label{eq:mdot1}
\end{equation}
where $\bar{\rho}(\xi)$ is the $\psi$-averaged density for a given impact
parameter $\xi$.  This average is obtained as:
\begin{equation}
\bar{\rho}\left(\xi\right)=
\frac{1}{2\pi}\int_0^{2\pi}\rho\left(\xi,\psi\right)\,\mathrm{d}\psi=
\rho_0\,\mathrm{e}^{-\xi^2/2h^2}\mathrm{I}_0\left(\frac{\xi^2}{2h^2}\right),
\label{eq:meanrho}
\end{equation}
where $\mathrm{I_n}(x)$ is the modified Bessel function of the first
kind. Substituting equation (\ref{eq:meanrho}) into equation
(\ref{eq:mdot1}) and performing the integral one obtains:
\begin{equation}
\dot{M}_{\mathrm{acc}}=4\pi\,\rho_0\,v_0\,
\xi_0^2\,\mathrm{e}^{-2\xi_0^2/h^2}\,
\left[\mathrm{I}_0\left(\frac{2\xi_0^2}{h^2}\right)+
      \mathrm{I}_1\left(\frac{2\xi_0^2}{h^2}\right) \right].
\label{eq:mdot2}
\end{equation}

The drag force felt by the point mass can be calculated considering
the net loss of momentum flux along the $x$-direction. Material coming into
the system from $x\rightarrow -\infty$ carries a $\rho v_0$ momentum flux,
while material leaving the system at $x\rightarrow \infty$ (fluid
parcels with $\xi\ge 2\xi_0$) takes a $\rho v_{\infty}$ momentum flux, thus
the drag can be obtained as 
\begin{equation}
F_{\mathrm{d}} = \dot{M}_{\mathrm{acc}}\,v_0+\int_{2\xi_0}^{\infty}
2\pi\,v_0\,\bar{\rho}(\xi)\,\left(v_0-v_{\infty}\right)\,\xi\,\mathrm{d}\xi.
\label{eq:Fd}
\end{equation}
The first term in the right-hand side is the drag caused by
accretion of material and the second term is a purely gravitational
drag by the material in the wake that will not be accreted.
In order to obtain a closed expression for $F_{\mathrm{d}}$, it is
convenient to introduce the new variables $l\equiv \xi/(2\xi_0)$, and
$\lambda\equiv \xi_0/h$. With these variables and replacing
Eqs.~(\ref{eq:vinf}) and (\ref{eq:meanrho}) into equation (\ref{eq:Fd}) the
drag force can be written as:
\begin{eqnarray}
F_{\mathrm{d}} &=& \dot{M}_{\mathrm{acc}}\,v_0 +
8\pi\rho_0\,v_0^2\,\xi_0^2\times \nonumber \\
& &
\left\{ \int_{1}^{\infty}\mathrm{e}^{-2\lambda^2\,l^2}
\mathrm{I_0}\left(2\lambda^2\,l^2\right)\left[l-\left(l^2-1\right)^{1/2}
\right]\mathrm{d}l\right\}.
\label{eq:Fd2}
\end{eqnarray}

The expressions for the mass accretion rate (equation \ref{eq:mdot2}), and
for the gravitational drag (equation \ref{eq:Fd2}) are easy to compute
numerically. 
Moreover, they can be further simplified in the
limiting cases of a thin and thick disk, as shown below.

\subsection{Thick disk limit: $\lambda \ll 1$}

If the scaleheight of the disk is much larger than the gravitational
radius (that is, $\lambda= (\xi_0/h) \ll 1$), 
the $\mathrm{I_0}$ Bessel function
in equation (\ref{eq:mdot2}) is close to unity, whereas
the $\mathrm{I_1}$ is close to zero. After substituting $\xi_{0}$ by
its definition in Equation (\ref{eq:xi0}), 
the accretion rate is approximately given by:
\begin{equation}
\dot{M}_{\mathrm{acc,thick}}\approx
\frac{4\pi\left(G\,M\right)^2\rho_0}{v_0^3}\,,
\label{eq:mdotthick}
\end{equation}
which is the same expression obtained for the uniform density case
\citep{2011MNRAS.418.1238C}.

The mass accretion rate can be written in terms of the disk surface
density given by
\begin{equation}
\Sigma = \int_{-\infty}^{\infty}\rho(y)\,\mathrm{d}y=
\int_{-\infty}^{\infty}\rho_0\,\mathrm{e}^{-y^2/h^2}\,
\mathrm{d}y=\sqrt{\pi}\,\rho_0\, h,
\label{eq:sigma}
\end{equation}
as~:
\begin{equation}
\dot{M}_{\mathrm{acc,thick}}\approx
\frac{4\sqrt{\pi} \left(G\,M\right)^2\Sigma}{h\,v_0^3}\,.
\label{eq:dotMthick}
\end{equation}

We may obtain an analytical expression for
the drag force by using the following approximations
\begin{subequations}
\begin{eqnarray}
\label{eq:thinasymp}
\mathrm{e}^{-2\,\tau^2}\,\mathrm{I_0}\left(2\,\tau^2\right)
 \approx & 1 &~;~2\,\tau^2 \ll 1, \\
\mathrm{e}^{-2\,\tau^2}\,\mathrm{I_n}\left(2\,\tau^2\right)
 \approx & \frac{1}{2\,\sqrt{\pi}\,\tau} &~;~2\,\tau^2 \gg
 1,
\label{eq:thickasymp}
\end{eqnarray}
\label{eq:asymp}
\end{subequations}
to split the integral in equation (\ref{eq:Fd2}) in two parts.
A first integral over $l$ in the range $1< l < l_0$ 
(such that $2\lambda^2\,l^2 < 1$),
and the second $x$-integral from $x_0$ to infinity
(such that  $2\lambda^2\,l^2> 1$),
where 
\begin{equation}
l_0=\frac{1}{2\sqrt{\pi}\,\lambda}
\label{eq:x0}
\end{equation}
is the junction of the asymptotic expressions in Eqs. (\ref{eq:asymp}).
More specifically,
\begin{eqnarray}
&&\int_{1}^{\infty}\mathrm{e}^{-2\lambda^2\,l^2}\mathrm{I_0}
\left(2\lambda^2\,l^2\right)\left[l-\left(l^2-1\right)^{1/2}
\right]\mathrm{d}l \approx\nonumber \\
&&\int_{1}^{l_0}\left[l-\left(l^2-1\right)^{1/2}
\right]\mathrm{d}l \nonumber \\
&&+\int_{l_0}^{\infty}\frac{l_0}{l}\left[l-\left(l^2-1\right)^{1/2}
\right]\mathrm{d}l,
\label{eq:intsplit}
\end{eqnarray}
Performing the integrals, and using the accretion rate of
equation (\ref{eq:mdotthick}), the total drag becomes
\begin{eqnarray}
F_{\mathrm{d,thick}}&\approx&4\pi\,\rho_0\,v_0^2\,\xi_0^2
\left[ -l_0^2 +  l_0\sqrt{l_0^2-1}   \right. \nonumber \\
&+&\left. 2\,l_0\,\arctan\left(\frac{1}{\sqrt{l_0^2-1}}\right )\right. \nonumber \\
&+&\left. \ln\left(l_0+\sqrt{l_0^2-1} \right) \right] .\label{eq:Fdthick1}
\end{eqnarray}
Remind that in terms of physical parameters, we have that 
$\xi_{0}^{2}=(GM)^{2}/v_{0}^{4}$ and $l_{0}=hv_{0}^{2}/(2\sqrt{\pi}GM)$.
The associated Coulomb logarithm, defined as 
$\ln \Lambda=F_{\mathrm d}/(4\pi \rho_{0}\xi_{0}^{2}v_{0}^{2})$,
corresponds to the term in square brackets in Eq.~(\ref{eq:Fdthick1}).
We see that the ambiguity in the definition of the minimum scale
of interaction that appears in linear theory (e.g., Ostriker 1999
for the homogeneous medium, or Rein 2012 for the case of a disk),
is removed in this approach. Moreover, no additional cut-off scale
needs to be introduced at large distances.
Using the hydrodynamical approach in the linear regime, 
Rein (2012) identified a Coulomb
logarithm but it is not the same as in the present work.

A yet simpler expression can be obtained taking the limit of
$l_0\gg1$ (i.e.~$\lambda \ll 1$; the thick disk limit)~:
\begin{equation}
F_{\mathrm{d,thick}}\approx
\frac{4\pi\,\rho_0\,(G\,M)^2}{v_0^2}
\left[\frac{3}{2}+\ln \left(\frac{h}{\sqrt{\pi}\xi_{0}}\right)
  \right].
\label{eq:Fdthick2}
\end{equation}
The error made by using the above expression instead of the exact
formula (Eq.~\ref{eq:Fd2}), is less than $2\%$ for $\lambda\leq 0.5$.
In terms of the disk surface density, the drag force in the thick
disk approximation can be expressed
as
\begin{equation}
F_{\mathrm{d,thick}}\approx
\frac{4\sqrt{\pi}\,\Sigma\,(G\,M)^2}{v_0^2 h}
\left[\ln \left(\frac{\mathrm{e}^{3/2}h}{\sqrt{\pi}\xi_{0}}\right)
  \right],
\label{eq:Fdthick3}
\end{equation}
where the factor $\mathrm{e}^{3/2}$ comes from inserting the term $3/2$ that
appears in the right-hand side of Equation (\ref{eq:Fdthick2}) in the argument
of the logarithm.

For a homogeneous medium, Cant\'o et al.~(2011) demonstrated that
the drag force for hypersonic bodies can be written as
\begin{equation}
F_{\mathrm{d}}\approx
\frac{4\pi\,\rho_0\,(G\,M)^2}{v_0^2} \ln \left(\frac{r_{\rm max}}{r_{\rm min}}\right)\,,
\label{eq:homog}
\end{equation}
where $r_{\rm max}$ is the largest impact parameter 
and $r_{\rm min}=\sqrt{\mathrm{e}}\xi_{0}/2$.
Comparing equations (\ref{eq:Fdthick3}) and (\ref{eq:homog}) we find that, 
in the case of a stratified medium, the ambiguity in the maximum
impact parameter is removed. Adopting $r_{\rm min}=\sqrt{\mathrm{e}}\xi_{0}/2$
and using Eq.~(\ref{eq:Fdthick3}),
we find $r_{\rm max}=\mathrm{e}^{2}h/(2\sqrt{\pi})\simeq 2.1h$ for a Gaussian
disk. Hence, this implies that the gravitational drag on a supersonic
body that is dropped suddenly at $t=0$ in a stratified
background, saturates asymptotically with time to a constant value.
This also holds true if the disk is infinitesimally thin (see the next
subsection and Muto et al.~2011).

\subsection{Thin disk limit: $ \lambda \gg 1$}

For a thin disk ($h\ll\xi_0$, or $~\lambda \gg 1$), one can use the
approximation of equation (\ref{eq:thinasymp}) in equation
(\ref{eq:mdot2}) to obtain the mass accretion rate:
\begin{equation}
\dot{M}_{\mathrm{acc,thin}}\approx 4\,\sqrt{\pi}\,\xi_0\,h \,\rho_0
\,v_0,
\label{eq:mdotthin1}
\end{equation}
or in terms of the disk surface density (equation \ref{eq:sigma}), after
substitution of equation (\ref{eq:xi0}), as
\begin{equation}
\dot{M}_{\mathrm{acc,thin}}\approx
4\,\xi_0\,v_0\,\Sigma=\frac{4\,G\,M\,\Sigma}{v_0}.
\label{eq:mdotthin2}
\end{equation}
Notice that the accretion rate has a much weaker dependence on the velocity
of the perturber $v_0$ than in the thick disk or uniform density
cases (Equation \ref{eq:dotMthick}), and that the dependence on the disk parameters is nicely folded
into a single parameter (i.e., the surface density).

The net gravitational drag (including the drag due to accretion)
on a body moving in a thin disk can be estimated combining
equations (\ref{eq:Fd2}, \ref{eq:thinasymp}, and \ref{eq:mdotthin1}),
yielding~:
\begin{equation}
F_{\mathrm{d,thin}}\approx 2\,\pi^{3/2}\,\rho_0\,v_0^2\,h\,\xi_0=
2\,\pi\,G\,M\,\Sigma.
\label{eq:Fdthin}
\end{equation}
Thus, the drag in a thin disk does not
depend on $v_0$ (i.e., the relative velocity between the
point mass and the surrounding environment), but solely on the
surface density $\Sigma$ of the disk and the mass $M$ of the
perturber.
Furthermore, the gravitational deceleration on the hypersonic perturber
$F_{\mathrm{d}}/M$ is independent of any of the parameters of the perturber
(i.e. its mass or velocity) and depends, only, on the surface mass
density of the disk.

It is common to compare the drag force in gaseous and collisionless
media. Following Binney \& Tremaine (1987), we have derived the dynamical
friction force in a collisionless medium but, instead of
assuming that the background particles are distributed homogeneously
in a three-dimensional medium, we assume that they 
are distributed in the $(x,z)$-plane, i.e.~two-dimensional geometry.
We find that, if the perturber moves at a velocity much larger than
the velocity dispersion of background particles, the dynamical
friction is also given by Eq.~(\ref{eq:Fdthin}).
We must stress here that the structure of the wake and the underlying
physics in gaseous media (where we may have gas accretion
onto the perturber) and in collisionless 
media (no accretion occurs) are very different.

\section{Numerical simulations}
\label{sec:num}

\subsection{The setup}
\label{sec:setup}

We have used the adaptive grid code {\sc Yguaz\'u-a}
\citep{2000RMxAA..36...67R} to perform a set of  numerical simulations to
be compared with the analytical model presented above. 
The version of the code employed solves the isothermal hydrodynamic
equations on a three-dimensional Cartesian grid. 
The equations are solved with a second order implementation of the
``flux vector splitting'' algorithm of \citet{1982LNP...170..507V}.

We use the same reference system as that described in \S \ref{sec:model},
in which the point mass is at rest at the center of the system. 
In units of $\xi_{0}$, the computational domain covers a region of
$[(-7.5,7.5),(0,7.5),(-7.5,7.5)]$, in $x$, $y$ and $z$,
respectively. The relevant scales that need to be solved
are $\xi_{0}$ and $h$, the gravitational radius and the scaleheight
of the disk, respectively. The domain is discretized in a five-level binary
adaptive grid with a maximum resolution equivalent to $512\times
256\times 512$ cells in a uniform mesh.
Simulations at half the resolution give similar results for the
accretion rate and the drag, but with a smoother structure of the
wake, due to numerical diffusion.
The gravitational attraction due to a point
mass centered at $(0,0,0)$ is added as an external source term. 
To avoid numerical artifacts a softening length of $10$ cells is
used for the gravitational force.
The accretion onto the point mass is emulated by artificially keeping
(at every timestep) a low density inside a hemisphere of radius
$0.15\,\xi_0$ ($\sim 5$ cells) centered around the point mass.

Different parameters have been used in the literature to measure
the degree of non-linearity ${\mathcal{A}}$. In the case of the 
Bondi-Hoyle accretion problem, we will use 
${\mathcal{A}}\equiv 2\xi_{0}/r_{s}$, which is the ratio between
the accretion radius ($2\xi_{0}$) and the softening radius $r_{s}$. 
In our simulations $r_{s}=0.3\xi_{0}$
and hence ${\mathcal{A}}\equiv 2\xi_{0}/r_{s} \simeq 6$. 
In astronomical systems,
${\mathcal{A}}$ may range 
between $1$ and $100$ for Earth-like and Jupiter-like planets,
up to much higher values for black holes.
However, a value of ${\mathcal{A}}\simeq 6$ is large enough to test
the success of our analytical formalism in a situation in which
non-linearity is important.

Since the problem has mirror symmetry on the $y=0$ plane, we only
simulate one half of the wake (for positive $y$), imposing a reflective
boundary condition at $y=0$. The subsequent analysis is done
considering the mirror symmetry.
The domain is initially filled with a streaming environment with a
density given by the Gaussian profile of equation
(\ref{eq:strat}), and a constant velocity $v_0$ in the $x$ direction.
An inflow condition (with the appropriate density profile and velocity)
is imposed on the $x=-7.5\xi_0$ boundary in order to replenish the streaming
environment. Outflow conditions are imposed on the remaining
boundaries. 

Because the flow is isothermal, the density gradient translates into a
vertical pressure gradient that tends to destroy the disk, which is
particularly problematic for the thin disk models (which have a larger
pressure gradient).
To avoid this we have added a vertical external force (in the $-y$
direction) that corresponds to the gravity force needed to maintain
hydrostatic equilibrium. Additionally, to avoid numerical problems due
to the large range of density values needed for the Gaussian
profile, we have set a lower limit beyond a
height of $y=3\,\xi_0$. Past this height the density is homogeneous
and the external vertical gravity is turned off. This has no noticeable
consequences on the results as most of the mass and momentum lie
closer to the midplane. 

\begin{deluxetable}{ccccc}
\tablecaption{Parameters of the simulations\tablenotemark{a}.
  \label{tab:models}}
\tablehead{
    \colhead{Model}
  & \colhead{$c_s$} 
  & \colhead{Mach number} 
  & \colhead{$h~[\xi_0]$}
  & \colhead{Regime}
}
\startdata
A  & $ 0.2$ & $5$ & $10^5$ & thick disk \\
B  & $ 0.2$ & $5$ & $2.0$ & thick disk \\
C  & $ 0.2$ & $5$ & $0.5$  & thin disk\\
D  & $ 0.1$ & $10$ &$0.5$  & thin disk
\enddata
\tablenotetext{a}{In all models $\rho_0=1$, and $v_0=1.0$.}
\end{deluxetable}

The parameters of the four models are summarized in Table
\ref{tab:models}. In all the models, we take units such 
that $\rho_{0}=1$ and $v_{0}=1$.
The first two simulations (A and B) correspond to a thick
disk. Given the large scaleheight used in model A, the ambient density is
close to constant inside the domain, thus the model is basically a 3D
version (albeit at a lower resolution due to computational
constraints) of the simulation in \citet{2011MNRAS.418.1238C}.
The other two models (C and D) correspond to a thin disk with the
same scaleheight but different sound speeds. 
Models C and D have Mach numbers of 5 and 10, respectively.
These two models were computed to test the result (of the analytic model) 
that the drag is independent of the Mach number of the flow. 

\subsection{Results}
\label{sec:results}
\begin{figure*}
\centering
\includegraphics[width=.85\textwidth]{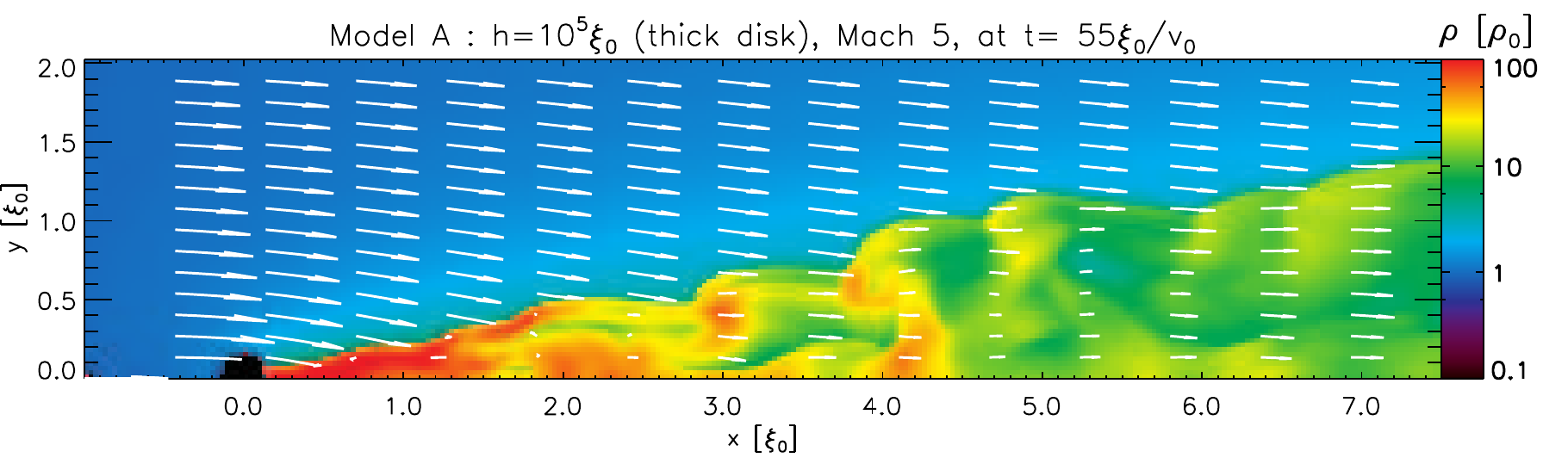}
\includegraphics[width=.85\textwidth]{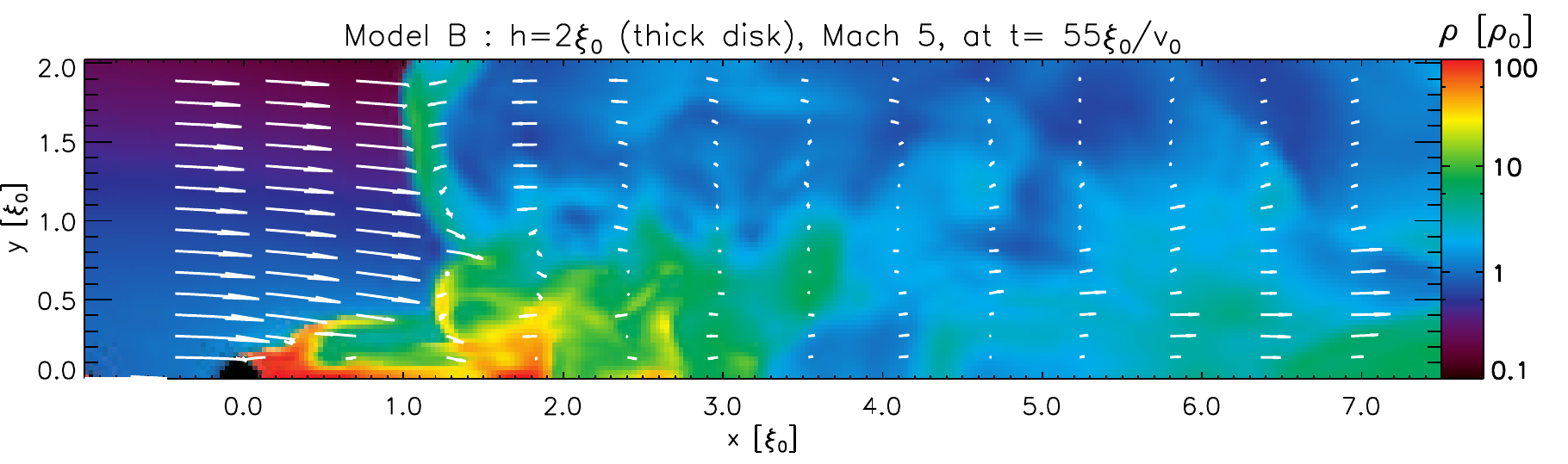}
\includegraphics[width=.85\textwidth]{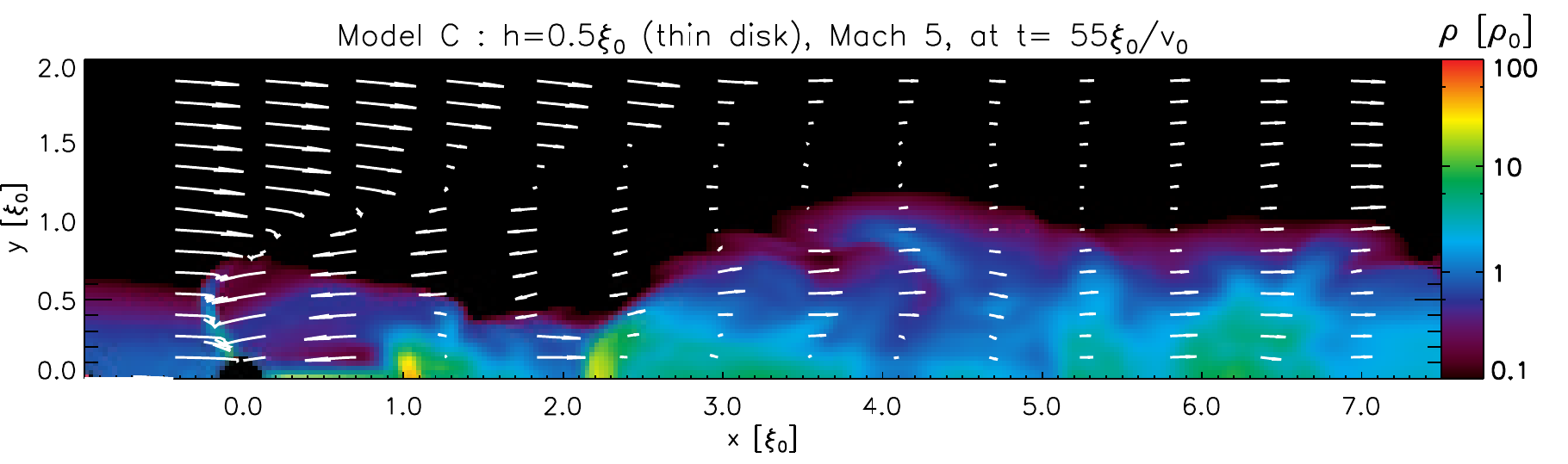}
\includegraphics[width=.85\textwidth]{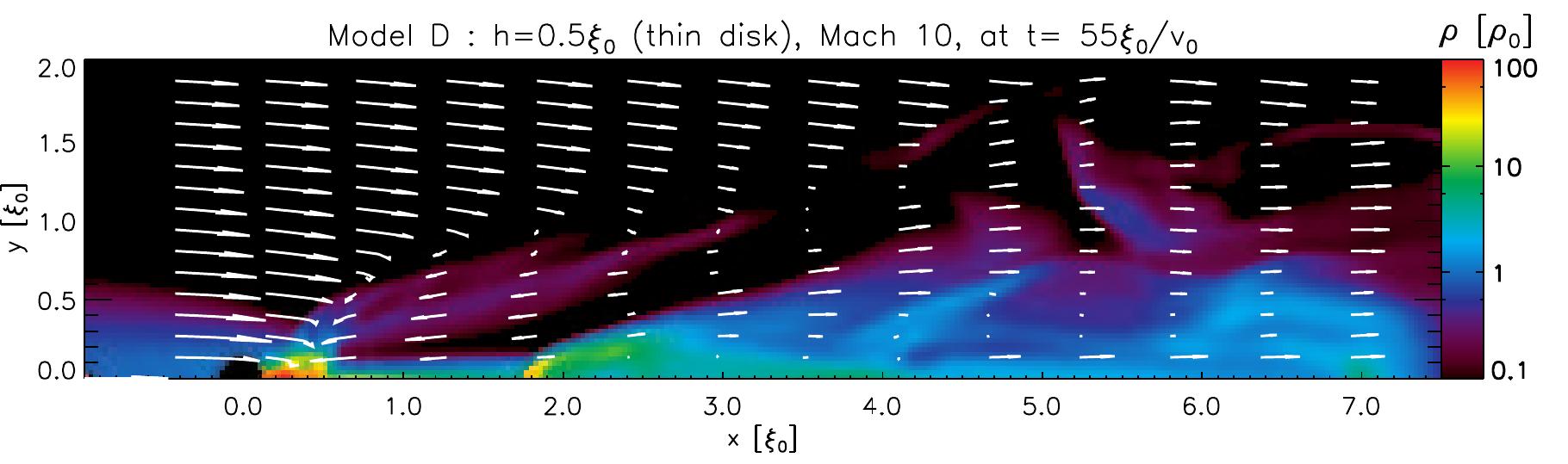}
\caption{Cuts of density and velocity along the $z=0$ plane 
  after an integration time
  of $t=55\xi_0/v_0$ for all the models (as indicated
  in the top labels). The gray-scale (color-scale in the online
  version) shows density, in units of the midplane density $\rho_0$, 
  in logarithmic scale 
  (as indicated by the bars on the right). 
  The arrows represent
  the velocity field. The panels only show an inner region of the
  entire computational domain where the wake is formed. The perturber
  is located at $(0,0)$ and the axes are given in units of $\xi_0$.}
\label{fig:cuts-4}
\end{figure*}
We allow the four models to run up to an evolutionary time of
${t=55\,\xi_0/v_0}$, ensuring that a quasi-stationary state is
achieved. Density maps for all the models after this integration time
are displayed in Figure \ref{fig:cuts-4}. In this figure we present a
zoom of a region close to the point mass. The snapshots show the
highly variable flow structure, formed downstream of the perturber centered
at $(0,0,0)$.
\begin{figure*}
\centering
\includegraphics[width=0.4\textwidth]{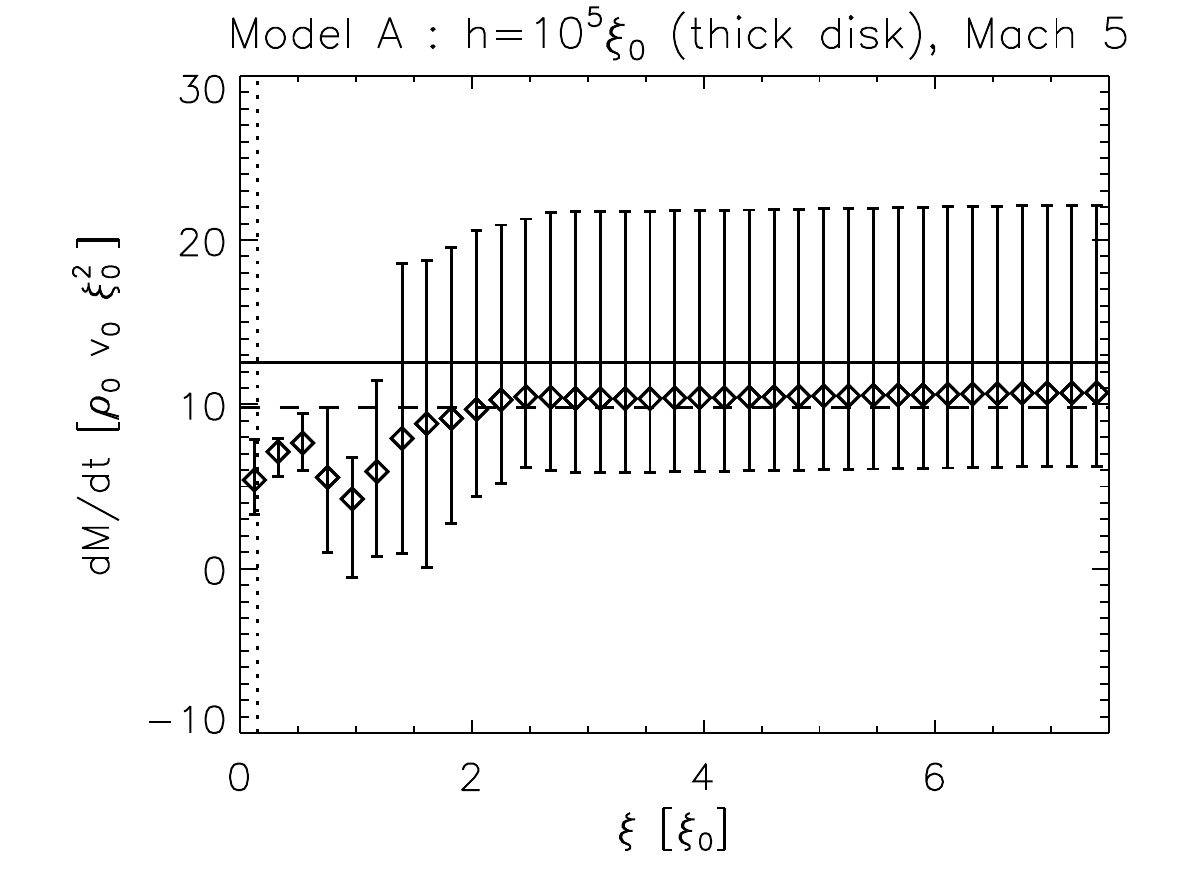}
\includegraphics[width=0.4\textwidth]{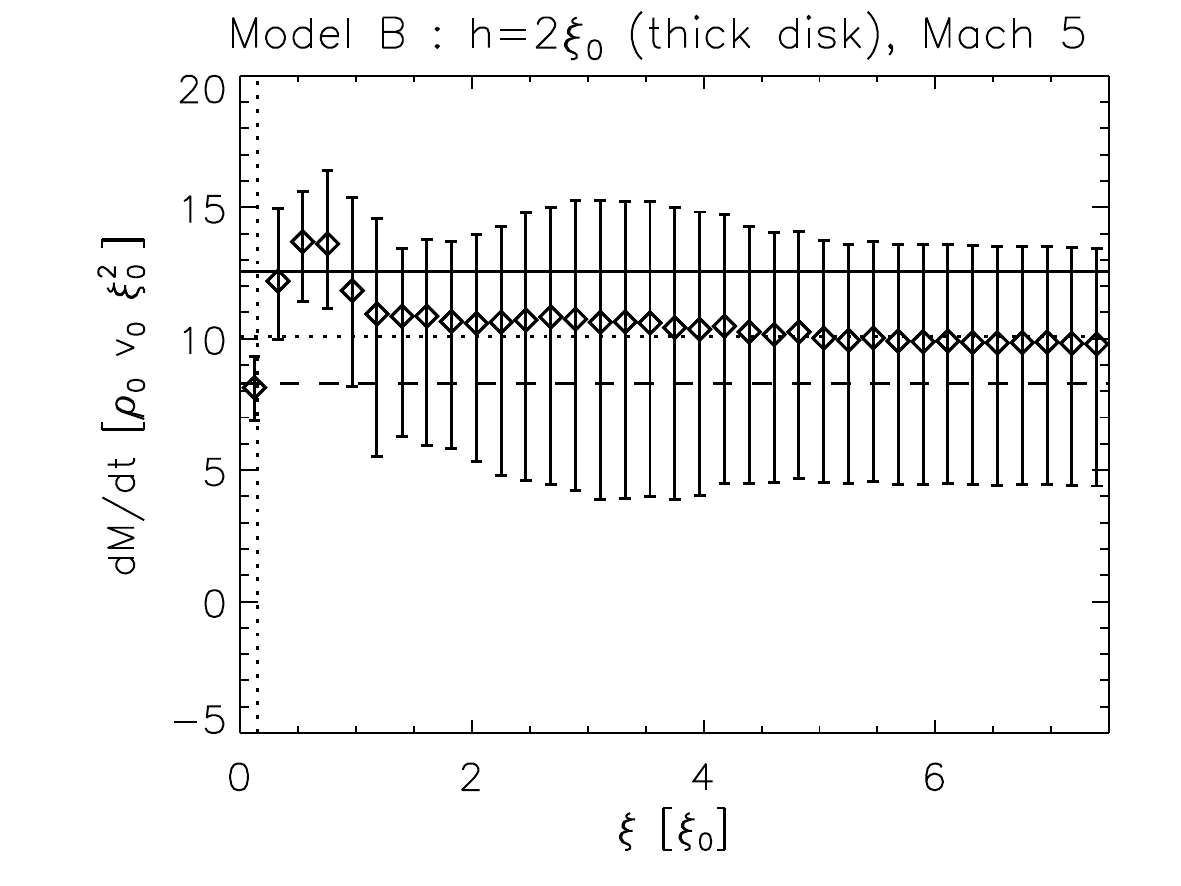}
\includegraphics[width=0.4\textwidth]{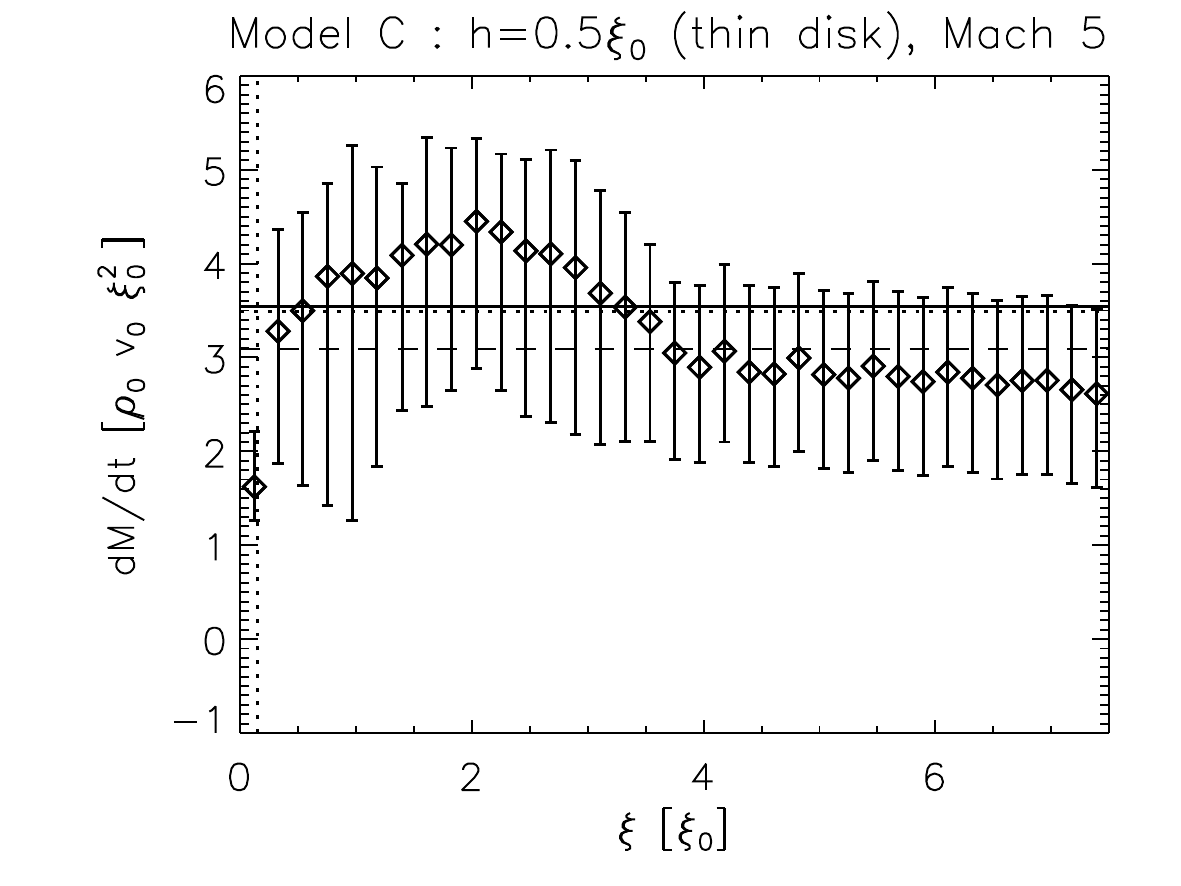}
\includegraphics[width=0.4\textwidth]{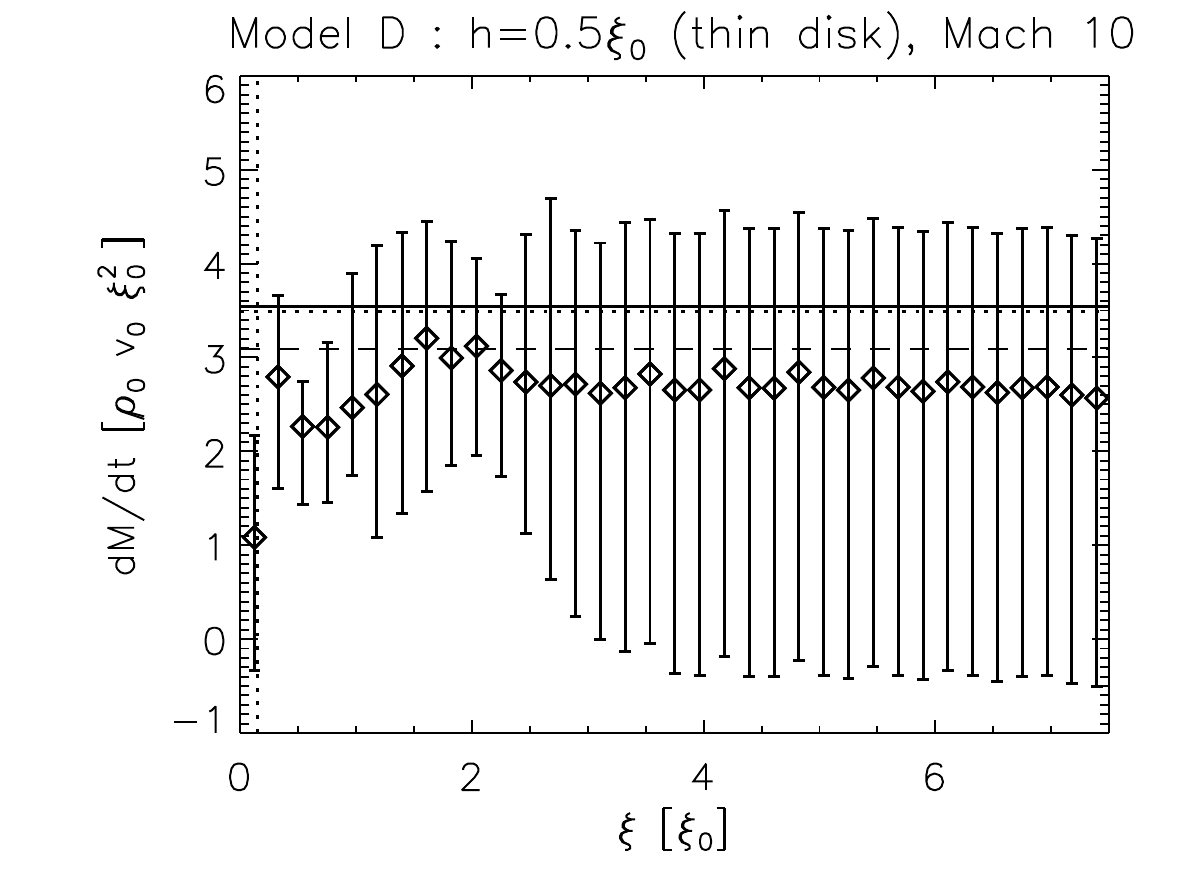}
\caption{Mass flux $\dot{M}$ (in units of $\rho_0\,v_0\,\xi_0^2$) 
  as a function of
  the cylindrical radius $\xi$ of the control volume, in units of
  $\xi_0$.   
  The symbols represent the
  average value (at each $\xi$) of the last five outputs in the simulations,
  corresponding to  evolutionary times from $51$ to
  $55\,\xi_0/v_0$. The error bars show the maximum departure from the
  mean value. The solid
  horizontal lines are the prediction from the analytical model:
  equation (\ref{eq:mdotthick}) for panels (a) and (b),  and equation
  (\ref{eq:mdotthin1}) for panels (c) and (d). The horizontal dotted
  line (overlaid with the solid line for A) shows the full solution
  from Eq. (\ref{eq:mdot2}).
 The dashed horizontal lines show the full solution, but including a
 correction due to the limited extent of the computational box (see
 text for details). The vertical dotted lines depict the size of the
 perturber.} 
\label{fig:mdot}
\end{figure*}

In order to compare the numerical simulations with the free-streaming
model, we have taken the density and velocity outputs and
computed the mass flux (per unit of time) $\dot{M}$ and the
net $x$-momentum flux,
over cylinders of varying radii ($\xi$). 
More specifically, 
\begin{equation}
\dot{M}(\xi)=\int_{S_{\xi}}\rho \vv\cdot d\SSS,
\end{equation}
\begin{equation}
F_{\mathrm D}(\xi)=\int_{S_{\xi}}\rho v_{x} \vv\cdot d\SSS,
\end{equation}
where $S_{\xi}$ is the surface of a cylinder with axis along the 
$x$-axis, radius $\xi$ and whose caps are placed at the boundaries 
of our domain,
at $x=-7.5\xi_{0}$ and at $x=7.5\xi_{0}$.
In a steady state, $\dot{M}$ should not depend on $\xi$ and must
coincide with the accretion of mass onto the perturber $\dot{M}_{\rm acc}$.
On the other hand, $F_{\rm D}(\xi)$ is the contribution of
the mass inside a cylinder of radius $\xi$ to the drag force. 
It includes both the force due to accretion of momentum
over the body surface and the gravitational force on the perturber
by its own wake. The drag force on the perturber $F_{\rm d}$ corresponds
to $F_{\rm D}$ when all the mass in the wake is included, that is,
in the limit $\xi\rightarrow \infty$.  
The results can be compared with the predictions of equations
(\ref{eq:mdotthick}), (\ref{eq:Fdthick2}), (\ref{eq:mdotthin2}) and
(\ref{eq:Fdthin}). 

The results for $\dot{M}$ as a function of the
cylindrical radius (i.e., the impact parameter) are presented in Figure
\ref{fig:mdot}. This figure gives the average
$\dot{M}$ from $t=51$ to $55$, in units of $\xi_0/v_0$, 
with error bars denoting the maximum variability
at each radius. The highly variable
mass flux, sign of a very turbulent wake, evidenced by the large error bars,
 is consistent with previous calculations \citep[see for
 instance][]{2011MNRAS.418.1238C}. 
The solid line in the figures correspond to the asymptotic
expressions for the thick (A, B) and thin (C, D) disk models, 
Equations (\ref{eq:mdotthick}) and (\ref{eq:mdotthin1}), respectively.
We have added the full solution for $\dot{M}$
(Equation \ref{eq:mdot2}), which is only appreciably different from
the solid line in Figure  \ref{fig:mdot}(b).
We can also see from Figure \ref{fig:mdot}
that $\dot{M}$ becomes more or less independent of the
impact parameter after $\xi=2\,\xi_0$, the impact parameter
beyond which the material remains unbound (see eq. \ref{eq:xi0}).

It is also noticeable that there is a systematic difference towards smaller
accretion rates compared to the analytical prediction. 
The reason for this offset  is the limited length of the computational
box. Material entering the downstream wake will travel
a distance \citep[see][]{2011MNRAS.418.1238C} 
\begin{equation}
x_m=\frac{2\xi_0}{\left(2\xi_0/\xi\right)^2-1}
\label{eq:xmax}
\end{equation}
along the positive $x$-axis before turning around and falling towards
the point mass. With the outflow boundary condition placed at $x=7.5\,\xi_0$,
all the mass that leaves the computational box is lost. From equation
(\ref{eq:xmax}) we can see that only material with $\xi\lesssim
1.77\,\xi_0$ is captured, while all the mass with an impact parameter
$\xi>1.77\,\xi_0$ leaves the domain and does not contribute to the
mass accretion. The effect of the limited domain size can be quantified by
integrating the mass accretion (equation \ref{eq:mdot1}) only up to
an impact parameter $1.77\,\xi_0$. The result is shown as a dashed
line in the plots of Figure \ref{fig:mdot}, agreeing well with the results from
the simulations.

\begin{figure*}
\centering
\includegraphics[width=0.4\textwidth]{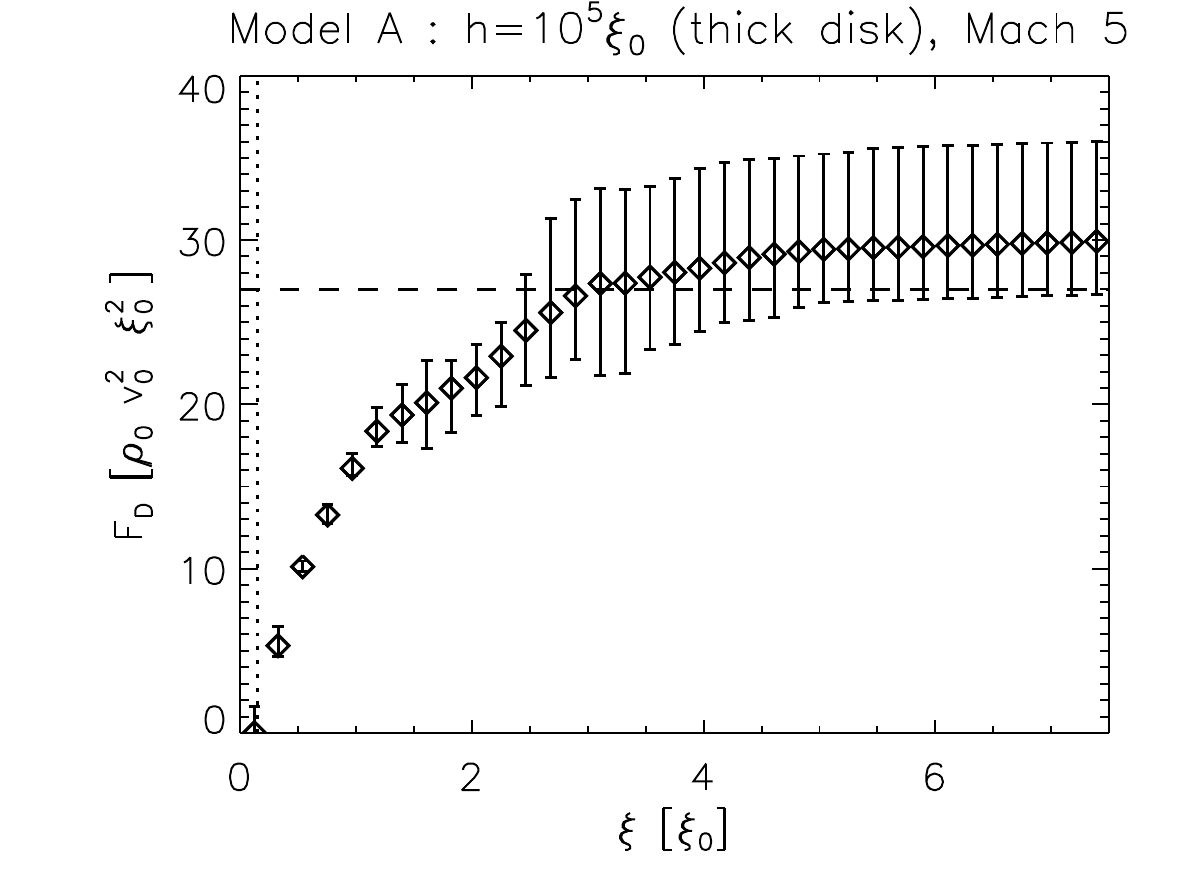}
\includegraphics[width=0.4\textwidth]{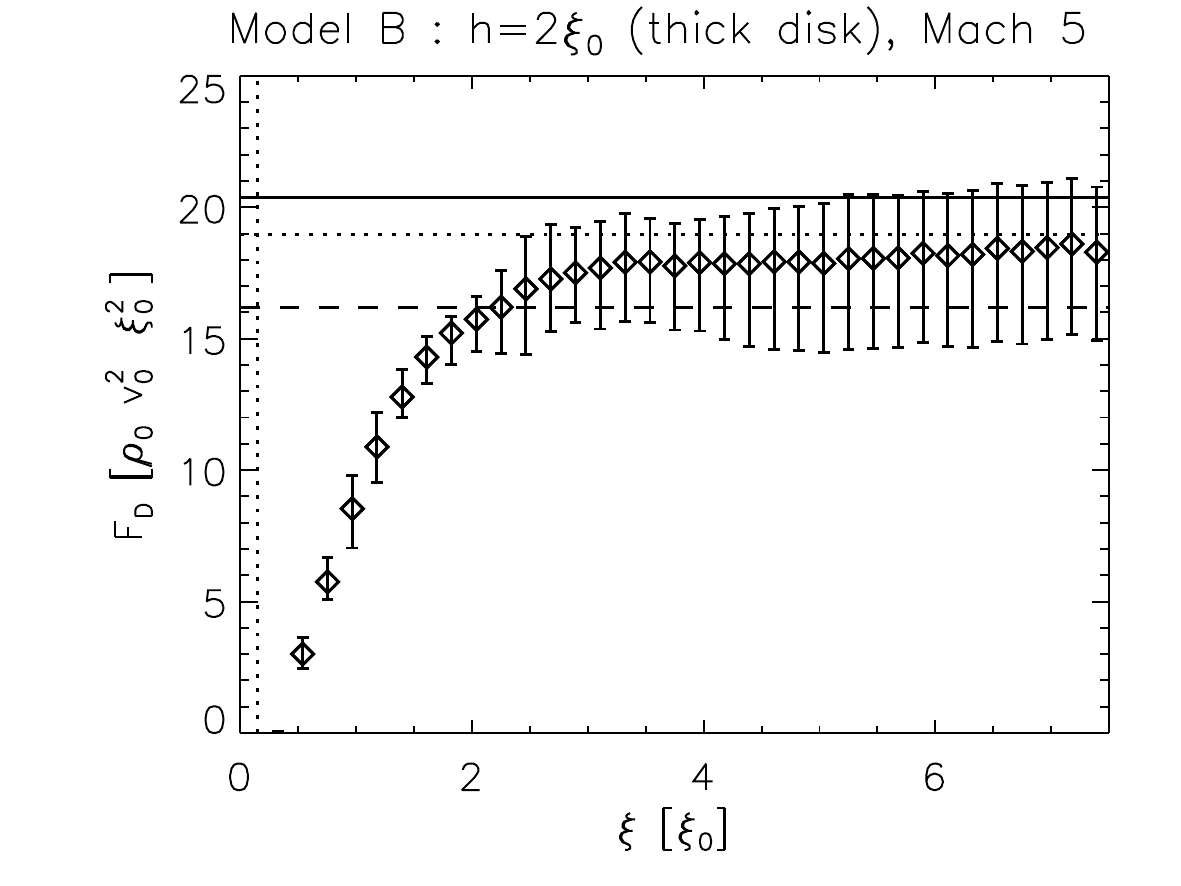}
\includegraphics[width=0.4\textwidth]{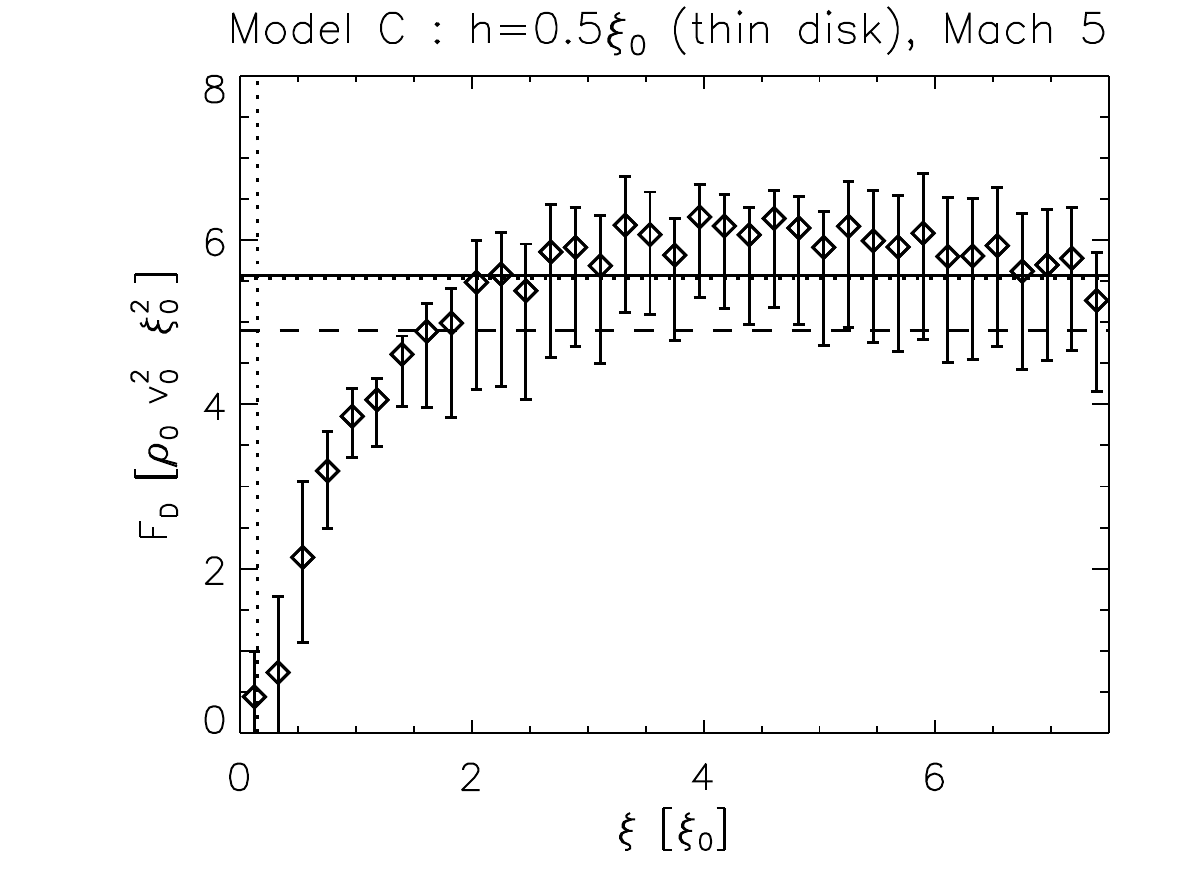}
\includegraphics[width=0.4\textwidth]{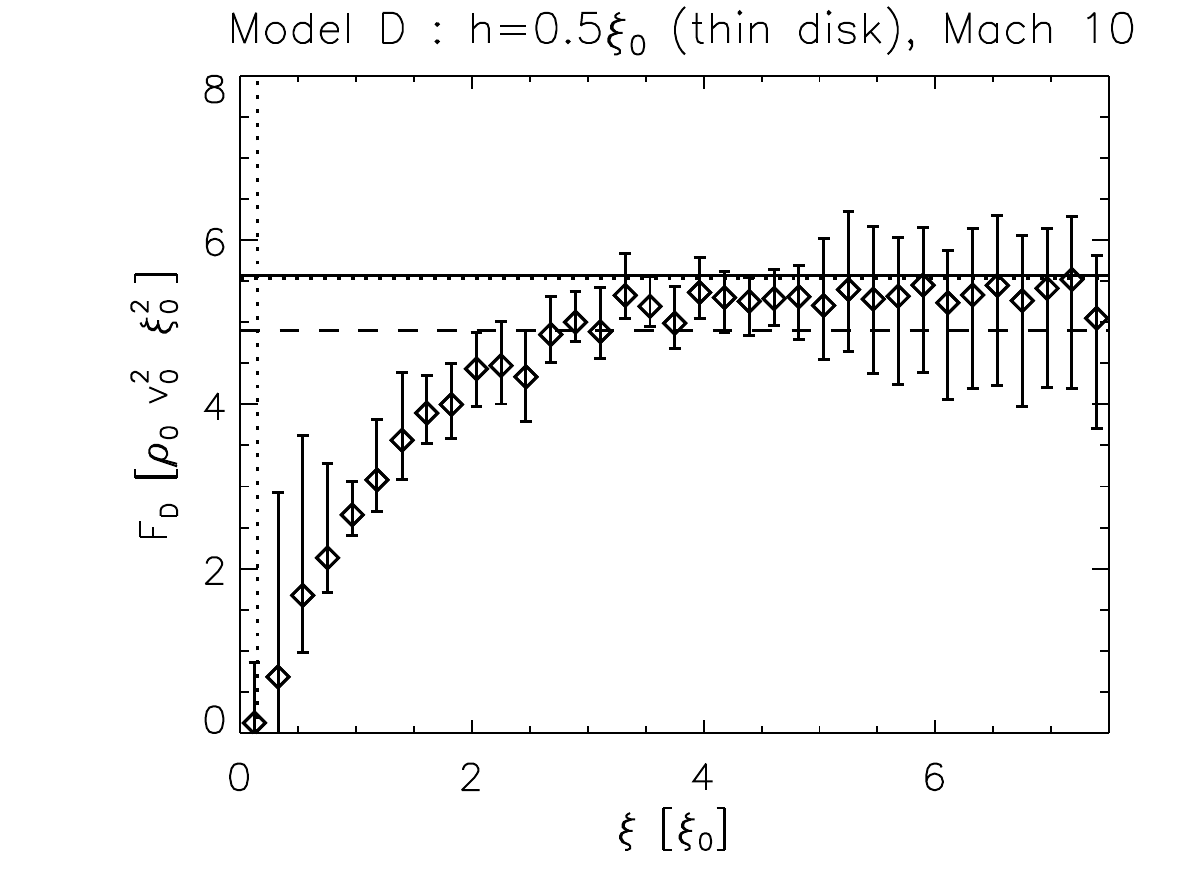}
\caption{$F_{\rm D}$ (in units of
  $\rho_0\,v_0^2\,\xi_0^2$) as a function of the cylindrical radius
  $\xi$ of the control volume, in units of $\xi_0$.
For large values of $\xi$,  it corresponds to the total drag onto the perturber.
The symbols represent the average value (at each $\xi$) of the last five outputs in
the simulations, corresponding to evolutionary times from $51$ to
$55\,\xi_0/v_0$, the error bars show the maximum departure from the
value. The solid horizontal
lines are the prediction from the analytical mode, equation
(\ref{eq:Fdthick2}) for A and B (For A the value is $\sim 156$,
which lies outside the range plotted, see text for more details) and
equation (\ref{eq:Fdthin}) for B and C. 
The horizontal dotted line (also $\sim 156$ for A) shows the
full solution from Eq. (\ref{eq:Fd2}).
The dashed lines include the additional correction for the finite
extent of the computational box, see text 
for details. The vertical dotted lines show the size of the perturber.}  
\label{fig:Fd}
\end{figure*}

The flux $F_{\mathrm{D}}$ as a function of impact
parameter (averaged over the same time window as the mass flux, see
above) is presented in Figure \ref{fig:Fd}.
The gravitational drag in our simulations also shows a
saturation for $\xi\geq 2\,\xi_0$, for all the models.

Comparing the two thin disk
models in Figs.~\ref{fig:Fd}(c,d) we can see that indeed
$F_{\mathrm{d}}$ is basically independent of the Mach number
(provided that the flow is hypersonic). 

The solid lines shown in Fig.~\ref{fig:Fd} correspond to the
analytical approximations given in equations (\ref{eq:Fdthick2})  and
(\ref{eq:Fdthin}),  for the thick and thin disk models,
respectively.
We have also included the full solution of Equation (\ref{eq:Fd2}) as
a dotted line, which overlaps the solid lines except for model B.
Since the net drag $F_{\rm d}$ has a term $\dot{M}_{\rm acc} v_{0}$ (see
Eq.~\ref{eq:Fd2}), we need to take into account that the accretion
rate is slightly smaller than predicted by the model because of the limited extent 
of the computational box in the $x$-direction. 
Including this effect, a new corrected value is derived, which we display 
as a dashed line in Figure \ref{fig:Fd}. 

However, for model A, which has a scaleheight ${h=10^5\,\xi_0}$, a
direct application of equations (\ref{eq:Fdthick2}) or (\ref{eq:Fd})
yields a drag force $F_{\mathrm{d}}\simeq 156~\rho_0\,v_0^2\,\xi_0^2$,
which is more than four times the  computed value.
The main reason for this discrepancy is that the model considers all
the mass in the disk, while the numerical simulation is restricted to
the drag by streamlines with impact parameter less than
$L_{y}=7.5\xi_{0}$.
In order to apply our analytical approach  to our simulation (or to a
truncated Gaussian disk), the integral in the right-hand side
of Equation (\ref{eq:Fd2}) should be performed between the
lower limit $l=1$ and the maximum impact parameter, in our case
$l_{\rm max}=L_{y}/(2\xi_{0})=3.25$.
For model A  it results in a drag force $F_{\mathrm{d}}\simeq
 27\,\rho_0\,v_0^2\xi_0^2$, which is the value plotted
 in Fig.~\ref{fig:Fd}(a) as a dashed line.
For the remaining models B, C and D, this correction is marginal
and the difference between the dashed, and the solid or dotted lines
is due to the correction in the accretion rate.

In general we see that the agreement between the analytical model  and
the numerical values is good.  The results for the  higher Mach number
model (D) agree slightly better, which is expected since the ballistic
model is applicable to hypersonic flows.

\section{Discussion: applicability of the model to a planet inside a Keplerian
  disk} 
\label{sec:kdisk}

The models presented in this paper (analytical and numerical) consider
a perturber traveling on a straight-line trajectory at constant
velocity within a medium with a Gaussian density stratification.
The medium has been regarded as a disk as one of the motivations was
to obtain the drag force on an object orbiting in such a system (e.g.
planets embedded in protoplanetary disks, or black holes in galactic nuclei).
In a real situation, the perturbers  move in complex orbits, the
density of the disk and the relative velocity of perturber with respect 
to the ambient gas change
with position, and the flow is not plane-parallel but presents differential
rotation (shear). 
One can use the `local' approximation, that is 
estimating the drag force at the
present location of the perturber ignoring
the curvature of the orbital motion, density gradients and shear effects,
as long as the curvature radius is
much larger than the scales involved in the model (i.e.~$\xi_0$ and $h$). 

The relevant scale for the drag force is 
$h$ in the case of a gravitational perturber moving in a thick disk. Hence, 
since the relevant scale for curvature effects and shear is
the size of the orbit, these effects can be ignored provided that
the aspect ratio of the disk is small (see also Rein 2012).

In the following we consider the case of a thin disk. 
Let us consider the example of a planet in orbit inside a Keplerian
disk to check under which conditions the model is adequate. 
A parcel of gas in a Keplerian disk around a star of mass $M_{*}$ will
have a velocity at a radius $r_0$ of \footnote{This is only valid if
 there is no radial pressure gradient.}~:
\begin{equation}
v_{\mathrm{k}}=\left(\frac{G M_{*}}{r_0} \right)^{1/2}.
\label{eq:vk}
\end{equation}

The relative velocity between the planet in orbit with a
velocity $\vv_{\mathrm{p}}$ and the disk is
$\vv_0=\vv_{\mathrm p}-\vv_{\mathrm{k}}$ and, therefore
\begin{equation}
v_{0}^{2}=v_{\mathrm p}^{2}+v_{\mathrm k}^{2}-2\vv_{\mathrm p}\cdot
\vv_{\mathrm k}=v_{\mathrm k}^{2}\left(\alpha^{2}+1-2\alpha\cos p\right),
\label{eq:vsquared}
\end{equation}
where 
\begin{equation}
\alpha= \frac{v_{\mathrm{p}}}{v_{\mathrm{k}}},
\end{equation}
and $p$ is the local pitch angle (the angle between $\vv_{p}$ and
the azimuthal direction). The planet is in corotation 
with the disk when $p=0$ and $\alpha=1$.

Combining equations (\ref{eq:xi0}), (\ref{eq:vk}) and (\ref{eq:vsquared}),
we can write the ratio between the
orbital radius and the planet gravitational radius as
\begin{equation}
\frac{r_0}{\xi_0} =\frac{M_*}{M_{\mathrm{p}}} \left(\alpha^2 +1 -2\alpha\cos p  \right),
\label{eq:xirat}
\end{equation}
where $M_{\mathrm{p}}$ is the mass of the planet.
Since the large-scale effects such as shear and curvature of both the orbit and
the disk will be important
when $r_0 \lesssim \xi_0$, 
we can use equation (\ref{eq:xirat}) to obtain the space of
parameters $\alpha$ and $p$ for which $\xi_{0}<r_0$ and, hence,
the local approximation is valid. This condition is
\begin{equation}
\frac{M_*}{M_{\mathrm{p}}} \left(\alpha^2 +1 -2\alpha\cos p  \right)>1.
\label{eq:xirat2}
\end{equation}
We see that when $p$ is close to $\pi/2$, the above condition is
fulfilled for any value of $\alpha$, provided that $M_{*}>M_{p}$.
The most stringent condition for $\alpha$ occurs when $p\approx 0$, 
i.e.~when the planet is close to pericenter
and apocenter. Taking $p\approx 0$, we find that  
the condition (\ref{eq:xirat2}) is not met for $\alpha$'s
in the interval between $\alpha_{-}$ to $\alpha_{+}$, 
where $\alpha_{\pm}\approx 1\pm \sqrt{M_{\mathrm p}/M_{*}}$.  
If we take for instance $M_{*}/M_{\mathrm{p}}=1000$ (similar to the
Sun/Jupiter system),  
it is simple to see that for orbits with eccentricities larger than $0.2$, 
it holds that $\alpha>\alpha_{+}$ at pericenter, whereas $\alpha<\alpha_{-}$ at
apocenter. Hence, we conclude that large-scale effects are small
at any radius as long as $e>0.2$.

A similar analysis can be done using the Hill radius of the planet in orbit~:
\begin{equation}
r_{\mathrm{H}}\approx a\,\left(1-e \right) \left(\frac{M_{\mathrm{p}}}{3M_*} \right)^{1/3},
\label{eq:hill}
\end{equation}
where $a$ is the semi-major axis of the planetary orbit, and $e$ its
eccentricity. The Hill radius roughly defines the region around the
planet in which its gravity dominates. Outside a
sphere with radius $r_{\mathrm{H}}$ the gravitational pull from the
central star is not negligible. Thus, in order to apply our model to the wake
of a planet in a disk, the Hill radius has to be larger than a few
times $\xi_0$ (the size of the wake).
Replacing equation (\ref{eq:xirat}) into (\ref{eq:hill}) we can write
\begin{equation}
\frac{r_{\mathrm{H}}}{\xi_0}=\frac{a\,\left(1-e \right) \left(\alpha^{2}+1
-2\alpha\cos p \right)} {3^{1/3}\,r_0}
\left(\frac{M_*}{M_{\mathrm{p}}} \right)^{2/3}.
\label{eq:hillrat}
\end{equation}
The gravitational pull from the star cannot be neglected if
$r_{\mathrm{H}}\lesssim \xi_0$. 
Again, the stringest constraint for $\alpha$ occurs when $p\simeq 0$.
In that case, the interval of $\alpha$ for which $r_{\mathrm{H}}<\xi_0$,
is $\alpha_{-}'<\alpha<\alpha_{+}'$,  where
\begin{equation}
\alpha_{\pm}'=1\pm 3^{1/6}\left[\frac{r_{0}}{a(1-e)}\right]^{1/2}\left(\frac{M_{\mathrm p}}{M_{*}}\right)^{1/3}.
\label{eq:alphacrit2}
\end{equation}
The factor in the square brackets in
Eq.~(\ref{eq:alphacrit2}), that is $r_{0}/(a(1-e))$,
varies between $1$ at periastron to
$(1+e)/(1-e)$ at apoastron. 
Let us consider the example of $M_{*}/M_{\mathrm{p}}=1000$
in a highly eccentric orbit with $e=0.8$. 
At the apoastron, we need $\alpha<\alpha'_{-}=0.64$ to have
$\xi_{0}<r_{\mathrm H}$. This condition for $\alpha$
is fulfilled because $v_{\mathrm p}/v_{\mathrm k}=\sqrt{(1-e)/(1+e)}$ 
at apoastron. 
In general, $\xi_{0}<r_{\mathrm H}$ at any radius of the orbit
provided that $e>0.2$.  
In conclusion, we have shown that, for
eccentric orbits with $e>0.2$ or retrograde disks, 
the curvature effects, as
well as the pull from the central star, on the wake can be neglected.

\section{Summary}
\label{sec:summary}

We present an analytical model for predicting the accretion rate and
gravitational drag on a point mass that travels hypersonically along
the midplane of a stratified medium with a Gaussian vertical
density profile.
The model considers that the trajectories of fluid parcels are
ballistic, and is a direct extension of the uniform density case
presented in \citet{2011MNRAS.418.1238C}.
The analytic model was then compared with a set of three-dimensional,
isothermal hydrodynamic simulations.
In contrast to the geometry used in \citet{2011MNRAS.418.1238C}, in
the present work cylindrical control volumes are used, which makes the
treatment easier and more natural given the symmetry of the problem.

The results can be summarized as follows~:

\begin{enumerate}[(i)]

\item Fully analytic expressions for the mass accretion rate
  $\dot{M}_{\mathrm{acc}}$ and  the gravitational drag force
  $F_{\mathrm{d}}$ are obtained assuming a free streaming ballistic
  flow.

\item Simpler expressions for $\dot{M}_{\mathrm{acc}}$ and
  $F_{\mathrm{d}}$ are obtained for two asymptotic limits; for a {\it
    thin disk}, in which the disk scaleheight is small compared to the
  gravitational radius, and for a {\it thick disk}, in which the disk
  scaleheight is large compared to the gravitational radius.  

\item In a thick Gaussian atmosphere with vertical scaleheight $h$,
  and surface density $\Sigma$ the drag force, including the 
contribution of the nonlinear part of the wake, is given by 
\begin{equation}
F_{\mathrm{d,thick}}\approx\frac{4\sqrt{\pi} (GM)^{2}\Sigma}{h\,v_{0}^{2}}\ln \left(\frac{r_{\rm max}}{r_{\rm min}}\right)\,,
\end{equation}
where $r_{\rm min}=\sqrt{\mathrm{e}}GM/(2v_{0}^{2})$ and 
$r_{\rm max}=\mathrm{e}^{2}h/(2\sqrt{\pi})$,
while the mass accretion rate is
\begin{equation}
\dot{M}_{\mathrm{acc,thick}}\approx\frac{4\sqrt{\pi} (GM)^{2}\Sigma}{h\,v_{0}^{3}}\,.
\end{equation}

\item We find that the mass accretion onto a point mass $M$ moving in a thin
  disk is proportional to $M$ and to the surface density $\Sigma$ of
  the disk,
\begin{equation}
\dot{M}_{\mathrm{acc,thin}}\approx\frac{4\,G\,M\,\Sigma}{v_{0}}\,.
\end{equation}

\item The gravitational drag force in the {\it thin disk}
  regime is independent of the flow velocity (provided that
  the flow is hypersonic), and is also proportional to $M\Sigma$.
\begin{equation}
F_{\mathrm{d,thin}}\approx 2\,\pi\,G\,M\,\Sigma\,.
\end{equation}
At the same time the gravitational deceleration of the hypersonic
perturber in a thin disk ($F_{\mathrm{d}}/M$) is independent of its
parameters (i.e. mass or velocity) and depends only on the surface
mass density of the disk. 

\item The mass accretion in the simulations was highly variable. The
  analytic prediction is slightly higher than the average values
  obtained from the simulations, but well within their dispersion.
  These differences might be due to
  the relatively low resolution of our present three-dimensional
  simulations.

\item The drag force computed from the simulations showed a good
  agreement with the analytic model, with a lower
  dispersion compared with the accretion rates. The result (from the
  analytic model) that, in a thin disk, the drag
  force is independent of the Mach number of the flow is confirmed
  by the numerical simulations.

\end{enumerate}

\acknowledgments
The referee is deeply thanked for an exhaustive scrutiny of the
manuscript.
This work has been supported by CONACYT grants 61547, 101356, 101975,
165584 and 167611, as well as DGAPA-UNAM IN105312 and IN106212 grants.


\begin{thebibliography}{4}
\expandafter\ifx\csname natexlab\endcsname\relax\def\natexlab#1{#1}\fi

\bibitem[Begelman et al.(1980)]{beg80}
Begelman, M. C., Blandford, R. D., Rees, M. J. 1980, Nature, 287, 307

\bibitem[Binney(1977)]{bin77}
Binney, J. 1977, MNRAS, 181, 735

\bibitem[Binney \& Tremaine(1987)]{bin87}
Binney, J. \& Tremaine, S. 1987, Galactic Dynamics, Princeton
University Press (Princeton, New Jersey) 

\bibitem[{{Bisnovatyi-Kogan} {et~al.}(1979){Bisnovatyi-Kogan}, {Kazhdan},
  {Klypin}, {Lutskii}, \& {Shakura}}]{1979SvA....23..201B}
{Bisnovatyi-Kogan}, G.~S., {Kazhdan}, Y.~M., {Klypin}, A.~A., {Lutskii}, A.~E.,
  \& {Shakura}, N.~I. 1979, \sovast, 23, 201

\bibitem[Bondi(1952)]{bon52}
Bondi, H. 1952, MNRAS, 112, 195 

\bibitem[Bondi \& Hoyle(1944)]{bon44}
Bondi, H., Hoyle, F. 1944, MNRAS, 104, 273

\bibitem[Bonnell et al.(2001a)]{bon01a}
Bonnell, I. A., Bate, M. R., Clarke, C. J., Pringle, J. E. 2001a, MNRAS,
323, 785

\bibitem[Bonnell et al.(2001b)]{bon01b}
Bonnell, I. A., Clarke, C. J., Bate, M. R., Pringle, J. E. 2001b, MNRAS,
324, 573 

\bibitem[{{Cant{\'o}} {et~al.}(2011){Cant{\'o}}, {Raga}, {Esquivel}, \&
  {S{\'a}nchez-Salcedo}}]{2011MNRAS.418.1238C}
{Cant{\'o}}, J., {Raga}, A.~C., {Esquivel}, A., \& {S{\'a}nchez-Salcedo}, F.~J.
  2011, \mnras, 418, 1238

\bibitem[Colpi et al.(1999)]{col99}
Colpi, M., Mayer, L., Governato, F. 1999, ApJ, 525, 720

\bibitem[Cuadra et al.(2009)]{cua09}
Cuadra, J., Armitage, P. J., Alexander, R. D., Begelman, M. C. 
2009, MNRAS, 393, 1423

\bibitem[Di Matteo et al.(2001)]{dim01}
Di Matteo, T., Carilli, C. L., Fabian, A. C. 2001, ApJ, 547, 131

\bibitem[Dokuchaev(1964)]{dok64}
Dokuchaev, V. P. 1964, Soviet Astron., 8, 23

\bibitem[Dotti et al.(2007)]{dot07}
Dotti M., Volonteri M., Perego A., Colpi M., Ruszkowski
M., Haardt F. 2010, MNRAS, 402, 682

\bibitem[Escala et al.(2005)]{esc05}
Escala, A., Larson, R. B., Coppi, P. S., Mardones, D. 2005, ApJ, 630, 152

\bibitem[Hoyle \& Lyttleton(1939)]{hoy39}
Hoyle, F., \& Lyttleton R. A. 1939, Proc.~Cambridge Philos.~Soc., 35, 405

\bibitem[Hoyle \& Lyttleton(1940a)]{hoy40a}
Hoyle, F., \& Lyttleton R. A. 1940a, Proc.~Cambridge Philos.~Soc., 36, 323

\bibitem[Hoyle \& Lyttleton(1940b)]{hoy40b}
Hoyle, F., \& Lyttleton R. A. 1940b, Proc.~Cambridge Philos.~Soc., 36, 325

\bibitem[Hoyle \& Lyttleton(1940c)]{hoy40c}
Hoyle, F., \& Lyttleton R. A. 1940c, Proc.~Cambridge Philos.~Soc., 36, 424

\bibitem[Just \& Kegel(1990)]{jus90}
Just, A., Kegel, W. H. 1990, A\&A, 232, 447

\bibitem[Just \& Pe\~narrubia(2005)]{jus05}
Just, A., Pe\~narrubia, J. 2005, A\&A, 431, 861

\bibitem[Khan et al.(2011)]{kha11}
Khan, F. M., Just, A., Merritt, D. 2011, ApJ, 732, 89

\bibitem[Kim \& Kim(2007)]{kimh07}
Kim, H., Kim, W.-T. 2007, ApJ, 665, 432


\bibitem[Klessen \& Burkert(2000)]{kle00}
Klessen, R. S., Burkert, A. 2000, ApJS, 128, 287

\bibitem[Klessen \& Burkert(2001)]{kle01}
Klessen, R. S., Burkert, A. 2000, ApJ, 549, 386

\bibitem[Maoz(1993)]{mao93}
Maoz E. 1993, MNRAS, 263, 75

\bibitem[Muto et al.(2011)]{mut11}
Muto, T., Takeuchi, T., Ida, S. 2011, ApJ, 737, 37

\bibitem[Namouni(2010)]{nam10}
Namouni, F. 2010, MNRAS, 401, 319

\bibitem[Nixon et al.(2011)]{nix11a}
Nixon, C. J., Cossins, P. J., King, A. R., Pringle, J. E. 2011a,
MNRAS, 412, 1591

\bibitem[Nixon et al.(2011)]{nix11b}
Nixon, C. J., King, A. R., Pringle, J. E. 2011b, MNRAS, 417, L66

\bibitem[Preto et al.(2011)]{pre11}
Preto, M., Berentzen, I., Berczik, P., Spurzem R. 2011, ApJ, 732, L26


\bibitem[{{Raga} {et~al.}(2000){Raga}, {Navarro-Gonz{\'a}lez}, \&
  {Villagr{\'a}n-Muniz}}]{2000RMxAA..36...67R}
{Raga}, A.~C., {Navarro-Gonz{\'a}lez}, R., \& {Villagr{\'a}n-Muniz}, M. 2000,
  Revista Mexicana de Astronomia y Astrofisica, 36, 67

\bibitem[Hein(2012)]{rein12}
Rein, H. 2012, MNRAS, 422, 3611

\bibitem[Roedig et al.(2012)]{roe12}
Roedig, C.,  Sesana, A. Dotti, M., Cuadra, J., Amaro-Seoane, P.,
Haardt, F. 2012, arXiv:1202.6063 

\bibitem[Ruderman \& Spiegel(1971)]{rud71}
Ruderman, M. A., Spiegel, E. A. 1971, ApJ, 165, 1

\bibitem[S\'anchez-Salcedo(2009)]{san09}
S\'anchez-Salcedo, F. J. 2009, MNRAS, 392, 1573


\bibitem[S\'anchez-Salcedo(2012)]{san12}
S\'anchez-Salcedo, F. J. 2012, ApJ, 745, 135 

\bibitem[S\'anchez-Salcedo \& Brandenburg(2001)]{san01}
S\'anchez-Salcedo, F. J., Brandenburg, A. 2001, MNRAS, 322, 67

\bibitem[Tanaka \& Haiman(2009)]{tan09}
Tanaka, T., Haiman, Z. 2009, ApJ, 696, 1798

\bibitem[Tremaine \& Weinberg(1984)]{tre84}
Tremaine, S., Weinberg, M. D. 1984, MNRAS, 209, 729

\bibitem[{{van Leer}(1982)}]{1982LNP...170..507V}
{van Leer}, B. 1982, in Lecture Notes in Physics, Berlin Springer Verlag, Vol.
  170, Numerical Methods in Fluid Dynamics, ed. E.~{Krause}, 507--512

\end{thebibliography}
\end{document}